\documentclass[%
superscriptaddress,
 amsmath,amssymb,
 aps, 
 prapplied,
 reprint,
]{revtex4-2}

\usepackage{graphicx}
\usepackage{dcolumn}
\usepackage{bm}
\usepackage{xcolor}

\begin{document}

\title{\textbf{An improved update rule for probabilistic computers} 
}%

\author{Andrew Rockovich}
 \email{Contact author: rockovich.6@osu.edu}
\affiliation{Department of Physics, The Ohio State University, Columbus, Ohio, USA}
\author{Gregory Lafyatis}
\affiliation{Department of Physics, The Ohio State University, Columbus, Ohio, USA}

\author{Daniel J. Gauthier}
\affiliation{ResCon Technologies, LLC. Columbus, Ohio, USA}

\date{\today}

\begin{abstract}
Many hard combinatorial problems can be mapped onto Ising models, which replicate the behavior of classical spins. Recent advances in probabilistic computers are characterized by parallelization and the introduction of novel hardware platforms. An interesting application of probabilistic computers is to operate them in `reverse' mode, where the network self-organizes its behavior to find the input bits that result in an output state.  This can be used, for example, as a factorizer of semiprimes. One issue with simulating probabilistic computers on standard logic devices, such as field-programmable gate arrays, is that the update rules for each spin involve many multiplications, evaluation of a hyperbolic tangent, and a high-resolution numerical comparison. We simplify these rules, which improves the spatial and temporal circuit complexity when simulating a probabilistic computer on a field-programmable gate array.  Applying our method to factorizing semiprimes, we achieve at least an order-of-magnitude reduction in the on-chip resources and the time-to-solution compared to recently reported methods.  For a 32-bit semiprime, we achieve an average factorization in $\sim$100 s.  Our approach will inspire new physical realizations of probabilistic computers because we relax some of their update-rule requirements. 
\end{abstract}

\maketitle

\section{Introduction\label{Sec:Intro}}

Modern digital computers have revolutionized nearly every aspect of society, leading to the information era.  Their success is based on computer hardware that implements a small set of deterministic logic gates for universal computing and a suite of algorithms that run on this platform.  Unfortunately, some important problems cannot be easily solved on a digital computer due to a lack of efficient algorithms \cite{Arora2009}.  By efficient, we refer to a problem that can be solved in a time that scales better than a super-polynomial function of its size and preferably as a low-order polynomial or better.  Hard problems include optimization tasks such as simulating quantum systems, factoring large semiprimes, the traveling salesman problem, binary satisfaction problems, etc.

The existence of hard problems motivates exploring other computing paradigms beyond the von Neumann architecture of digital computers. One promising approach is to use quantum mechanical spins to represent information, which can take advantage of quantum superposition and entanglement \cite{Nielsen2020}.  This concept gained great interest when algorithms that can solve a subset of hard `classical' problems efficiently were identified on a quantum computer \cite{Shor1997,Grover1996}.  As with digital computers, a few discrete quantum gates can be composed to realize universal quantum computation.  Considerable world-wide effort is underway to achieve fault-tolerant quantum computers \cite{Xanadu, Google, Zuchongzhi2025}.

\begin{figure*}[htb]
    \centering
    \includegraphics[width=0.95\linewidth]{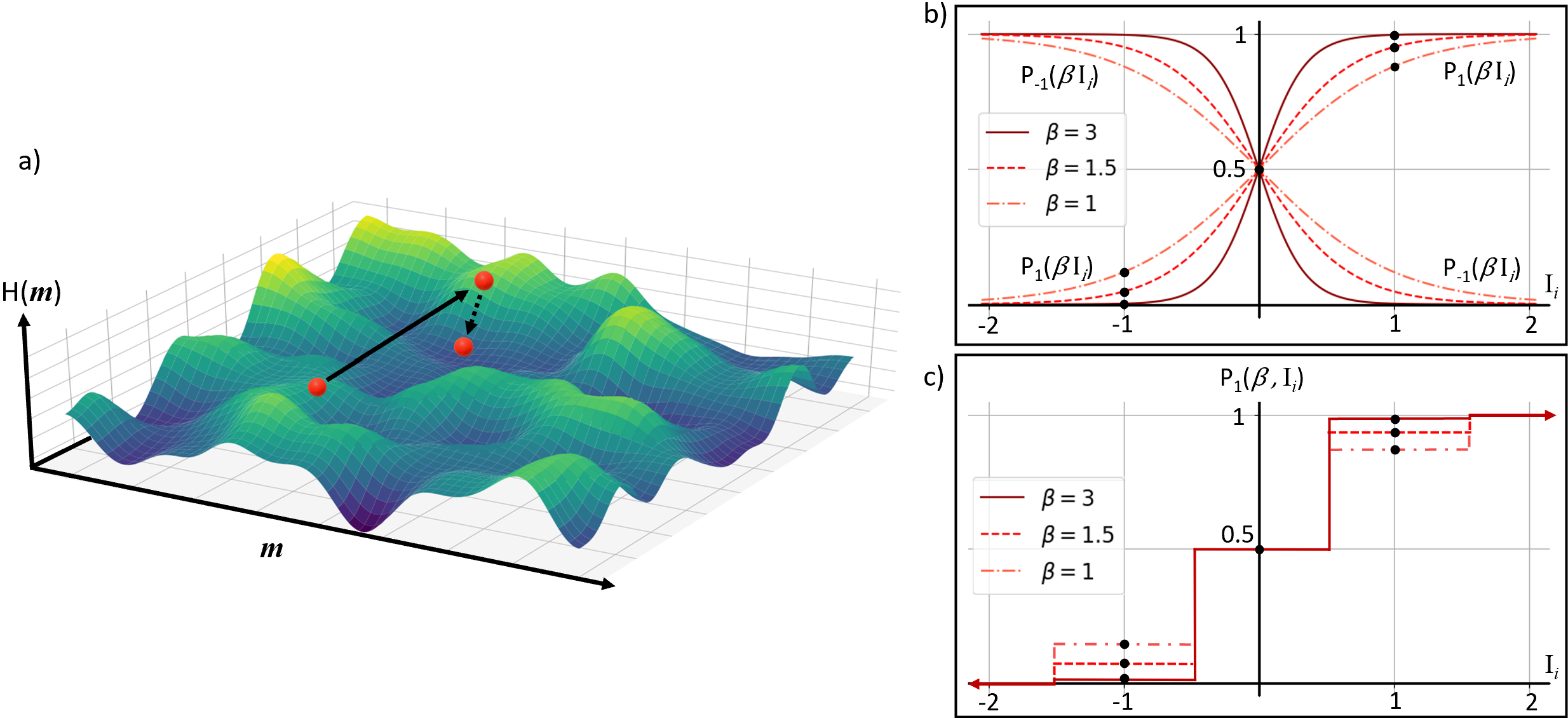}
    \caption{\textbf{P-Bit network finds an energy minima.} a) A simplified illustration of the energy landscape, where the horizontal dimensions are a flattened N-dimensional $\bm{m}$. The red balls are the energies of given states, the solid line is an energetically unfavorable spin flip, and the dashed line is an energetically favorable flip. b) Typical update rules' continuous spin-flip probabilities. c) Our discretized and truncated spin-flip probabilities.}
    \label{fig:storyboard}
\end{figure*}

A parallel effort is underway to determine whether networks of emulated atoms interacting via their magnetic dipole moments -- classical spins -- can offer an advantage for solving hard problems \cite{Kirkpatrick-Annealing, BoltzmannMachine1984, Fu1986, Mohseni2022}.  They are typically called Ising machines and are studied using the theory of statistical mechanics. Problems are encoded in the interactions between the networked spins, and the solution corresponds to the energy ground state of the network.  A properly designed network self-organizes its dynamics to seek out the ground state, as illustrated in Fig.~\ref{fig:storyboard}a. As with classical digital and quantum computers, a small number of gates can be composed to realize universal computing.  

Ising machines are characterized by a temperature that perturbs the spin orientation. These perturbations help the system find the global ground state and prevent it from being stuck in a local energy minima.  Typically, the system starts at a higher temperature and is cooled slowly in a process known as annealing, analogous to optimization algorithms running on classical computers such as simulated annealing \cite{Kirkpatrick-Annealing}. \textcolor{black}{More sophisticated optimization routines include parallel tempering \cite{Earl2005} and simulated bifurcation \cite{Hayato2019, Hayato2021}}. A variety of devices, such as photonic oscillators \cite{Yamamoto2017}, magnetic tunnel junctions \cite{Sutton2020,Borders2019,Si2024}, and memristors \cite{Woo2022}, are being developed to realize scalable Ising machines.  Because of their inherent noisy behavior, these machines are sometimes referred to as probabilistic computers, which we adopt here, and the fundamental carrier of information is a probabilistic bit or P-Bit.

One interesting aspect of P-Bit networks is that they can operate in both the forward and reverse modes.  In the forward mode, the P-Bits corresponding to the `input' bits of the system are clamped at the input values, and the `output' P-Bits of the network self-organize to provide the output solution. This is the standard operation mode of digital logic circuits.  In the reverse mode, the output P-Bits are clamped, and the network self-organizes so that the input P-Bits take on values consistent with the output states.  The reverse mode of a probabilistic multiplier circuit can be used, for example, to find the factors of a semiprime, where the output P-Bits encode the semiprime, and the network is driven to its ground state with the input P-Bits encoding the two factors.

The primary purpose of our paper is to demonstrate that the P-Bit update algorithm for a probabilistic computer can be greatly simplified using the method illustrated in Figs.~\ref{fig:storyboard}b) and c).  \textcolor{black}{We approximate the typical sigmoidal probabilistic update rules by a step function over integer input values. Then, we truncate the update rules and sparsify the network \cite{Aadit2022}. This enables us to eliminate all computationally expensive clocked arithmetic that is typically required for P-Bit updates, converting them to a much more efficient look-up-table format.} This result greatly speeds up the network dynamics and reduces the hardware resources needed to emulate a probabilistic computer on complementary metal oxide semiconductor (CMOS) devices. It may also suggest other P-Bit hardware solutions.

In the next section, we introduce the well-established two-body interaction-based P-Bit update rules and show how a P-Bit network can be composed to solve hard problems.  We introduce our new update rule in Sec.~\ref{Sec:Innovation}.  We apply our approach to operate the fundamental AND and full-adder gates in reverse mode in Sec.~\ref{Sec:LogicGates}.  We emulate a P-Bit network using a field-programmable gate array and demonstrate high-accuracy operation with fewer resources than previous work.  We then show how to realize a multiplier operating in reverse mode, which can be used for semiprime factorization.  We demonstrate high-accuracy factorization of 32-bit numbers (16-bit $\times$ 16-bit factors) in Sec.~\ref{Sec:multipliers}.  As reported recently by other groups, we observe that the time-to-solution for this problem scales exponentially with the number of bits.  In Sec.~\ref{Sec:Conclusions}, we compare our results with the current state of the field and look toward the future.

\section{Background \label{Subsec:Background}}

The lattice of spins is constructed out of ferromagnetic (antiferromagnetic) interacting adjacent P-Bits (state $m_i$ in the bi-polar range $\{-1,1\}$) via positive (negative) interaction and described by the Hamiltonian
\begin{equation}\label{Eq:H}
    H(\bm{m}) = -\sum_ih_im_i-\sum_{i<j}J_{ij}m_im_j. 
\end{equation}
Here, the weights form a symmetric ($N\times N$) matrix $\textbf{J}$ associated with quadratic interactions, and $N$ is the system size (total number of P-Bits). In addition, each P-Bit may be biased toward spin `up' (1) or `down' (-1) described by the ($N\times 1$) column vector $\textbf{h}$.  

For a network at a pseudo-inverse-temperature $\beta$, Eq.~(\ref{Eq:H}) yields the familiar Boltzmann probabilities 
\begin{equation}\label{Eq:probs}
    P(\bm{m}) = \frac{\textrm{exp}[-\beta H(\bm{m})]}{\sum_\zeta\textrm{exp}[-\beta H(\bm{m}_{\zeta})]},
\end{equation}
where the denominator is the partition function, a sum over all possible state combinations. As $\beta\rightarrow \infty$ (pseudo-temperature $\rightarrow 0$), only the P-Bit configurations with the lowest energy will have a non-zero probability.  To find the global energy minima, $\beta$ varies in time.  It starts `hot' initially (say, $\beta=1$), and then increases toward $\infty$ (zero pseudo-temperature) with a predefined waveform \cite{Kirkpatrick-Annealing}.

The interaction weights can be designed directly using linear programming \cite{Onizawa2021_DesignFramework}.  The constraints to the linear program are: entries appearing in the circuit truth table take on the lowest energy, and those that are not in the truth table have a larger energy. It is common to use integer-constrained linear programming so that all elements of $\mathbf{J}$ and $\mathbf{h}$ are integers.   For larger circuits, $\textbf{J}$ and $\textbf{h}$ can be found by adding those of the smaller circuits.  With these interactions, the P-Bits network self-organizes to find the solution to the problem: The configurations that appear in the circuit truth table.

\textcolor{black}{Recent efforts show that sparse Ising machines scale better with problem size than those with all-to-all connectivity \cite{Sajeeb2025,Nikhar2024}. Additional sparsification provides further potential benefits by reducing the number of conflicting assignments \cite{Aadit2022,Sajeeb2025}.} This allows for simultaneous updating of large blocks of P-Bits, greatly speeding up the energy minimization process.  Sparsification is done by adding auxiliary COPY gates to the circuit, which we illustrate in Appendix~\ref{Appendix:Multiplier}.  Identifying simultaneous update blocks is done using a greedy graph-coloring algorithm. The different color blocks are updated sequentially using phase-shifted clocks so that every P-Bit is updated in a single clock cycle.

The Hamiltonian (\ref{Eq:H}) specifies the energy landscape conceptually illustrated in Fig.~\ref{fig:storyboard}a).  A `particle' on this energy landscape will seek to lower its energy by moving in the direction of the negative gradient.  Typically, the P-Bits are updated one at a time, familiar from stochastic gradient descent algorithms \cite{Ruder2016}, to prevent simultaneous flips of adjacent P-Bits that create a contradiction.  For updating one at a time, the update rule for each P-Bit is found from 
\begin{equation}\label{Eq:I}
    I_i=-\frac{\partial H}{\partial m_i}=\sum_jJ_{ij}m_j+h_i,
\end{equation}
where $I_i$ is the directivity, the propensity to make P-Bit $i$ equal to 1 (up, $I_i>0$) or -1 (down, $I_i<0$) on the update step.  

When the network is at a finite pseudo-temperature, the spins are perturbed so that the update rule is not exact: There is a chance that the update will cause the P-Bit to move in an energy-increasing direction.  As mentioned above, these perturbations are needed to prevent the system from getting stuck at a local minimum.  In this case, the update rule is usually written as \cite{camsari2017}
\begin{equation}\label{Eq:m}
    m_i=\textrm{sgn}[\textrm{tanh}(\beta I_i)-\textrm{rand}(-1,1)],
\end{equation}
where the hyperbolic tangent confines the directivity to the range (-1,1), 
\begin{equation}
\textrm{sgn(x)} =
    \begin{cases} 
      -1 & x < 0 \\
      0 & x = 0 \\
      1 & x>0  
   \end{cases},
\end{equation}
and rand is a uniformly-distributed random number in the range (-1,1) representing the perturbation.  

When $\textrm{tanh}(\beta I_i)=0$, which is when $I_i=0$ independent of $\beta$, the P-Bit state is scrambled by the rand function. For $|\beta I_i|>>1$, the hyperbolic tangent function is saturated, and the rand function has almost no effect on the state.  Overall, the probability that the P-Bit attains the state $\pm$1 is given by
\begin{equation}\label{eq:update1_prob}
P(m_i=\pm 1) = \frac{1}{2}[1\pm\tanh(\beta I_i)].
\end{equation}
Figure~\ref{fig:storyboard}b) shows the conditional probabilities as a function of $I_i$ for different $\beta$.

\begin{figure}
    \centering
    \includegraphics[width=\linewidth]{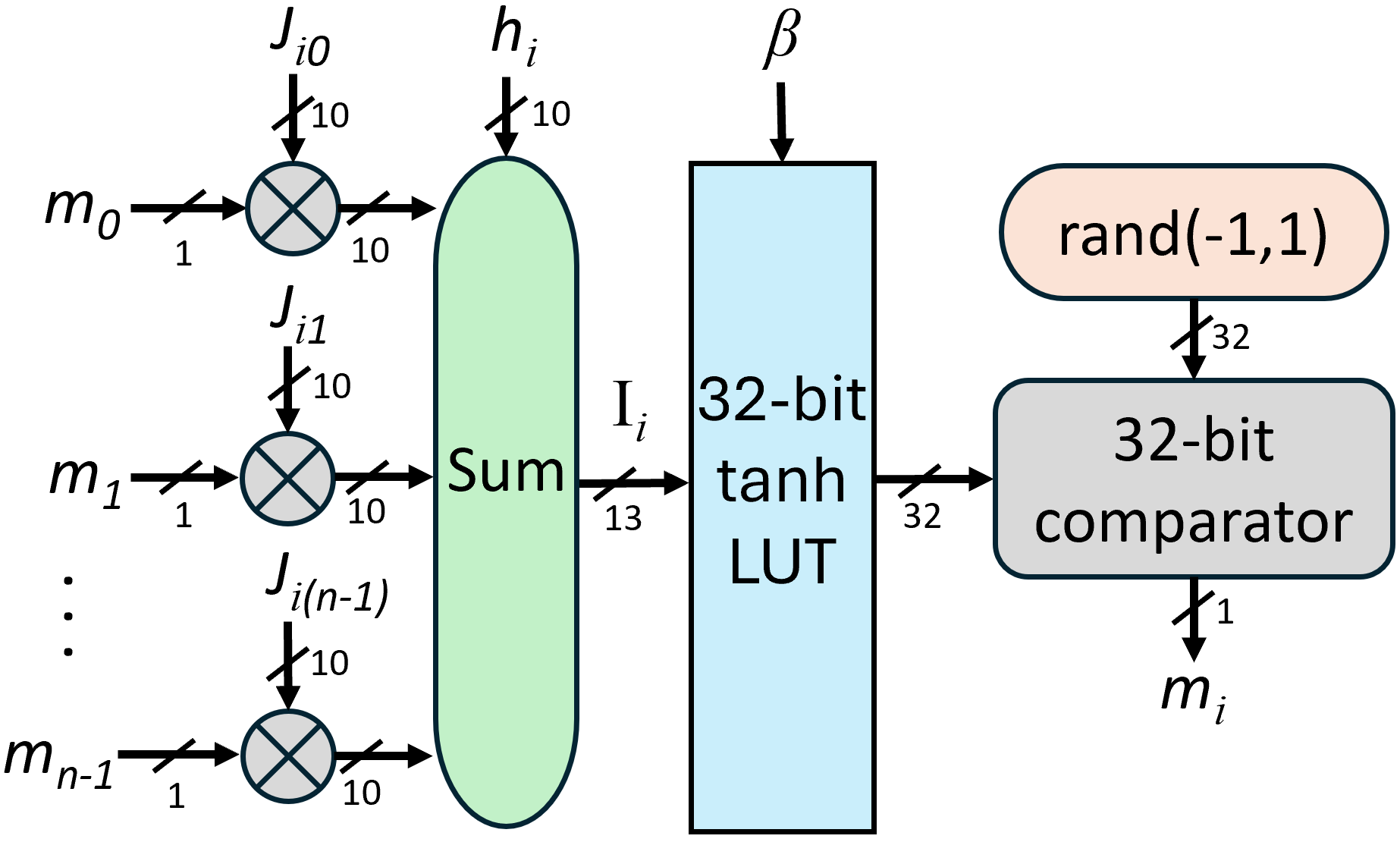}
    \caption{\textbf{Previous P-Bit simulation circuits.} Typical circuitry required on an FPGA to update the $i$th P-Bit. The gray circles are multipliers, and the bit-widths of each value are given (adapted from \cite{Aadit2022}).}
    \label{fig:typicalCircuit}
\end{figure}

In recent realizations of P-Bit networks on CMOS-based field-programmable gate arrays (FPGAs) \cite{Pervaiz2017,Aadit2022}, Eq.~(\ref{Eq:m}) for each P-Bit is evaluated as faithfully as possible, with a typical circuit shown in Fig.~\ref{fig:typicalCircuit}.  Here, the state of each P-Bit input to P-Bit $i$ is multiplied by the interaction weights using a finite-precision  multiplier, the values are added and combined with the bias, and the hyperbolic function is applied by a $\beta$-dependent 32-bit look-up-table (LUT).  The result is compared to a 32-bit pseudo-random number generator that spans the range (-1,1), which outputs the P-Bit state (either -1 or 1).  

This circuit requires fast and slow clocks.  It takes multiple fast clock cycles to execute the P-Bit circuit and slower clocks to perform the sequential updates of the P-Bit in the network.  Furthermore, the circuit consumes substantial on-chip resources.  In the next section, we describe methods to drastically reduce its complexity and speed up its operation.

\section{An Efficient P-Bit Update Rule\label{Sec:Innovation}}

There are several steps we take to simplify the update rule~(\ref{Eq:m}).  The first is to rewrite the update rule as
\begin{equation}
m_i = \mathrm{RNG}(b) \mathrm{sgn}(I_i),
\end{equation}
where the \textit{biased random number generator} is defined by
\begin{equation}
\mathrm{RNG}(b) =
 \begin{cases}
 +1~~\mathrm{with~probability}~b \\
 -1~~\mathrm{with~probability}~\textcolor{black}{(1-b)},
 \end{cases}
\end{equation}
and the bias is in the range (0.5,1) and given by 
\begin{equation}\label{eq:B}
b = \frac{1+|\mathrm{tanh}(\beta I_i)|}{2},
\end{equation}
where $b=1/2$ for an unbiased random number generator and $b=1$ for a fully biased bit (that is, $m_i= \mathrm{sgn}(I_i)$).  

With these definitions, the P-Bit update is governed by the sign function in Eq.~(\ref{Eq:m}) independent of $\beta$, and a bit-flip governed by the bias, which depends on $\beta$. We show below that decomposing the update rule in this way greatly reduces the P-Bit circuit complexity.

The second step is to realize that $I_i$ is an integer when the components of $\mathbf{J}$ and $\mathbf{h}$ are integers.  Most values of $I_i$ saturate the hyperbolic tangent function in Eq.~(\ref{eq:B}) when $\beta$ is not too small (say, $\beta\geq1$). This means that the hyperbolic tangent function only needs to be evaluated for a few of the smaller values of $I_i$ indicated by the dots in Fig.~\ref{fig:storyboard}b).  To stress this point, we redraw the probabilities as step functions in Fig.~\ref{fig:storyboard}c).  We then need to evaluate \textcolor{black}{$b$} for a few values of $\beta$ to achieve fast convergence to the ground state of the P-Bit network.

The final step is to realize that sgn($I_i$) can be found using a LUT, which matches the architecture of FPGAs typically used to simulate P-Bit networks.  For P-Bit $i$ with $M$ connections, there are $2^M$ entries in the LUT, which becomes impractical when $M$ is large.  However, the sparsification of $\mathbf{J}$ greatly helps in this regard, which is already done to allow for block updating of the network.  In particular, we follow Aadit \textit{et al.} \cite{Aadit2022} and set $M\leq5$.  Another advantage of this approach is that the LUT only compares binary values as opposed to the 32-bit comparator and the LUT used for the high-precision hyperbolic tangent in the circuit shown in Fig.~\ref{fig:typicalCircuit}.

Continuing with the analysis of our approach, the probability of an energy-increasing P-Bit assignment is only $\sim 0.5\%$ when $I_i=\pm2$ and $\beta\sim1.5$.  Based on this observation, we make a further simplification for the bias
\begin{equation}\label{eq:Bapprox}
b =
 \begin{cases}
  (1+|\mathrm{tanh}(\beta I_i)|)/2~~~&|I_i| \leq 1 \\
  1~~&|I_i|>1.
 \end{cases}
\end{equation}
We find that this provides enough randomness to allow the network to escape local energy minima.

It is customary to translate the bipolar spin states in the range $\{-1,1\}$ to binary states in the range $\{0,1\}$ because this range matches the traditional description of CMOS logic gates.  We translate the P-Bits states as \cite{Aadit2022}
\begin{equation}
s_i = \frac{m_i+1}{2},
\end{equation}
and the weight matrices as
\begin{eqnarray}
\mathbf{J}_{binary} = 2\mathbf{J} \\
\mathbf{h}_{binary} = \mathbf{h} - \mathbf{J}\mathcal{A},
\end{eqnarray}
where $\mathcal{A}$ is a $(N\times 1)$ vector of 1's.  For the rest of the paper, we use the binary convention and drop the \textit{binary} sub-script.

The remaining part of our new update rule is to describe an efficient method for the biased random number generation.  The goal is to realize a sequence that is mostly 0 (binary range) with probability \textcolor{black}{$b$} and 1 with probability \textcolor{black}{$1-b$}.  This can be accomplished by performing the logical $R$-input AND operation on $R$ independent and unbiased RNG's.  The probability of obtaining a 1 is then
\begin{equation} \label{eq:b_final}
    \textcolor{black}{b =
 \begin{cases}
  1/2~~&I_i=0 \\
  1/2^R~~~&|I_i| = 1 \\
  1~~&|I_i|>1.
 \end{cases}}
\end{equation}
We find that choosing one value of $R$ is enough to obtain fast convergence to the ground state; that is, no annealing is needed.

In our implementation below, we feed the $R$-input AND gate with randomly selected bits from 46-bit linear feedback shift-registers (LFSRs), which is a pseudo-random number generator with a repeat pattern every $2^{46}$ bits.  This is much larger than the number of P-Bit update steps needed for our experiments. Each color requires a set number of LFSRs, each contributing 46 random bits, depending on the number of P-Bits in that color, whether they contain cases of $I_i=\pm1$ or $I_i=0$, and R. The maximum number of LFSRs we require for a single color is for a color in the 16-bit $\times$ 16-bit multiplier in Sec.~\ref{subsec:largerMults} with 608 P-Bits, requiring 39 46-bit LFSRs. We select $R$ at random (without replacement) and AND them to generate one biased bit. Unbiased bits are drawn from the same set without replacement.

\section{Individual Logic Gates\label{Sec:LogicGates}}

For small probabilistic circuits, we find \textbf{J} and \textbf{h} directly with linear programming as described in Sec.~\ref{Subsec:Background}, without auxiliary bits or sparsification. We consider the AND and Full Adder (FA) gates because they are the components of a binary multiplier described in the next section.  We also only show experimental results for the reverse mode of the P-Bit network because the forward mode can be found using standard logic. The experiments are conducted on a Xilinx Zynq Ultrascale+ RFSoC 4$\times$2 FPGA (XCZU48DR-2FFVG1517E) with 930k system logic cells, and 425k configurable logic blocks (CLBs) consisting of 6-input LUTs and flip-flops (FFs).  Data is collected from the P-Bit network using the on-chip integrated logic analyzer (ILA) and transferred to a laptop computer via the JTAG interface for off-line analysis.  

\subsection{AND\label{subsec:AND}}

\begin{figure*}
    \centering
    \includegraphics[width=0.7\linewidth]{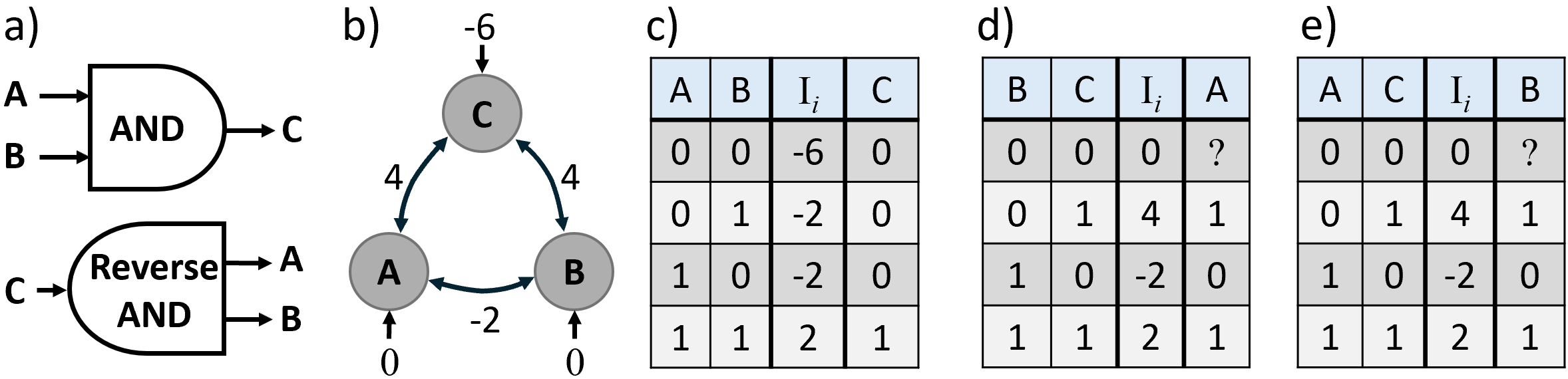}
    \caption{\textbf{AND gate.} a) Typical circuit depiction. b) Probabilistic formulation, with weight and bias terms in binary form. LUT for updating output P-Bit c) $C$, d) $A$, where the ambiguity (?) corresponds to equal probability of $A=0$ and $A=1$, and e) $B$, which is symmetric to d).}
    \label{fig:AND}
\end{figure*}

\begin{figure}
    \centering
    \includegraphics[width=\linewidth]{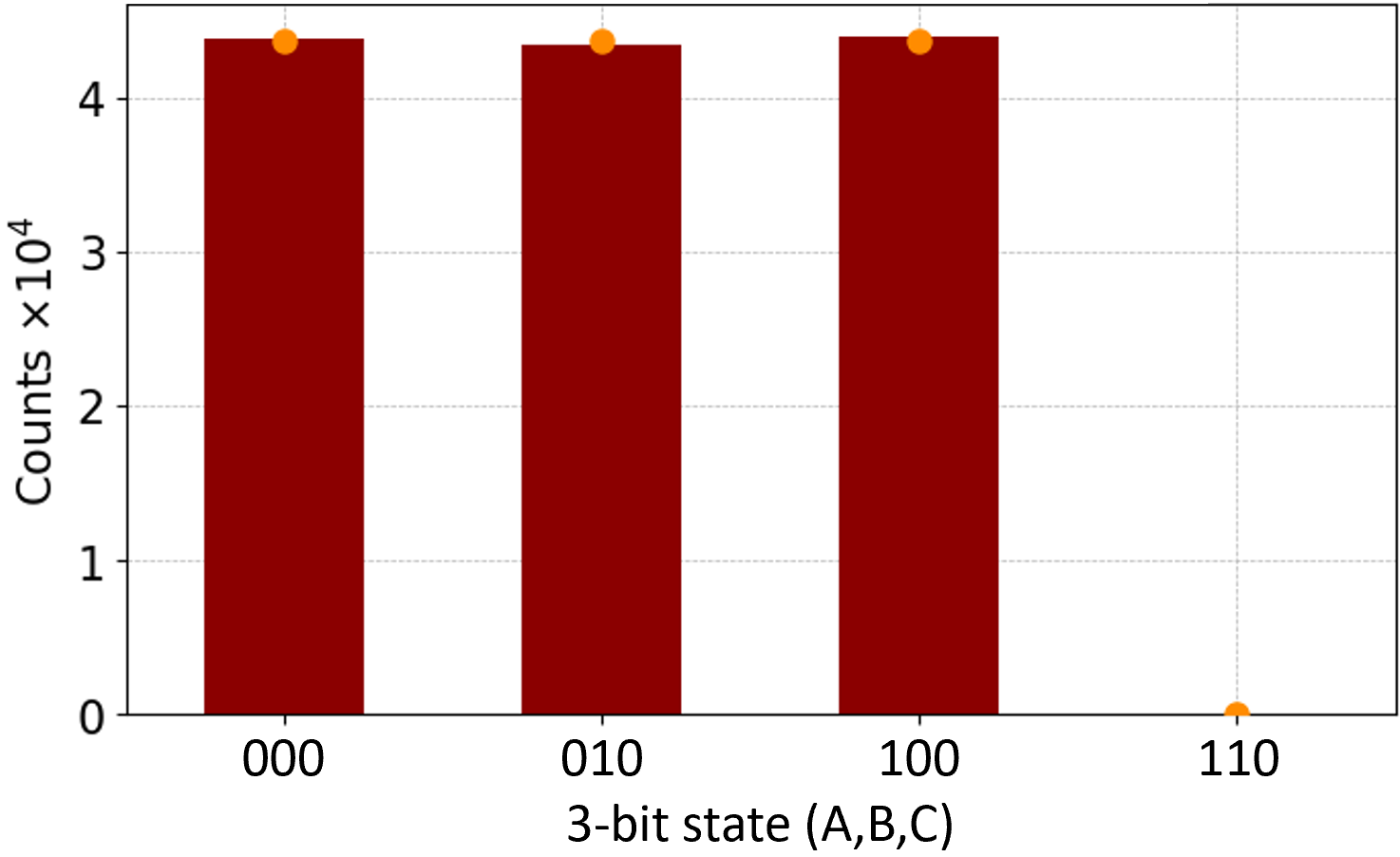}
    \caption{\textbf{AND gate experiments.}  Frequency diagram when running the AND gate in reverse mode, with P-Bit $C$ clamped to 0, sampled after each of the 131,072 full system updates.  The orange dots show the Bolztmann probabilities given by Eq.~(\ref{Eq:probs}).}
    \label{fig:ANDdata}
\end{figure}

The probabilistic AND gate is a 3-P-Bit logic gate illustrated in forward and reverse modes in Fig.~\ref{fig:AND}a), and the weights and biases are given in Fig.~\ref{fig:AND}b). When operating the AND gate in forward mode, the input P-Bits A and B are clamped to the desired input values, and the output P-Bit C changes its value to find the minimum energy of the network.  Randomness is never needed for the forward AND gate: $|I_i|>1$ in every line of the LUT seen in Fig.~\ref{fig:AND}c), so the updates to C are deterministic.

The situation is more complex when the probabilistic AND gate is operated in reverse mode.  There are two cases to consider: $C=0$ and $C=1$. When $C=1$, the only possible state consistent with the AND truth table is $A=1$ and $B=1$.  Figures~\ref{fig:AND}d) and e) show the update rules for P-Bit $A$ and $B$, respectively, when $\beta \rightarrow \infty$.  We see that $I_i$ takes on large values for $C=1$, and hence we do not apply randomness using the approximate update rule \textcolor{black}{(\ref{eq:b_final})}.

When $C=0$, we have three possible input configurations: $\{A,B\} = \{0,0\}$, $\{0,1\}$ and $\{1,0\}$, which all have the same (ground state) energy and should be equally likely.  When updating $A$, the update is deterministic if $\{B,C\} = \{1,0\}$ for $I_i=-2$.  On the other hand, $A$ ($B$) is set to an unbiased random number when $\{B,C\} = \{0,0\}$ ($\{A,C\} = \{0,0\}$) because $I_i=0$.

We update $A$ and $B$ sequentially at 400 MHz on the positive edges of a single clock with two outputs shifted by $120^{\circ}$.  Once $A$ and $B$ update, we measure them on the positive edge of a third clock output with an additional $120^{\circ}$ phase shift.  Thus, full updating of the network and reading out its state happens in a single clock cycle.  We start the system in state $\{A,B\} = \{0,0\}$. When started in a higher energy state, the first updated P-Bit puts the system into an allowed state after the first P-Bit is updated. 

Figure~\ref{fig:ANDdata} shows the experimental results for the probabilistic AND gate operating in reverse mode with $C=0$.  We see that the network visits all three states with near equal probability, as expected based on our discussion above regarding the Boltzmann probabilities given by Eq.~(\ref{Eq:probs}).

Based on the clock frequency, we perform 8.0$\times10^8$ P-Bit flips per second (FPS). The ILA is likely the limiting factor for the clock frequency, as this speed is around the usual maximum before timing violations. The on-chip resources for the AND P-Bit network are 10 LUTs and 14 FFs, not including the ILA used for data collection, which consumes an additional 1036 LUTs and 1827 FFs.

\subsection{Full Adder \label{subsec:FullAdder}}

\begin{figure*}
    \centering
    \includegraphics[width=0.75\linewidth]{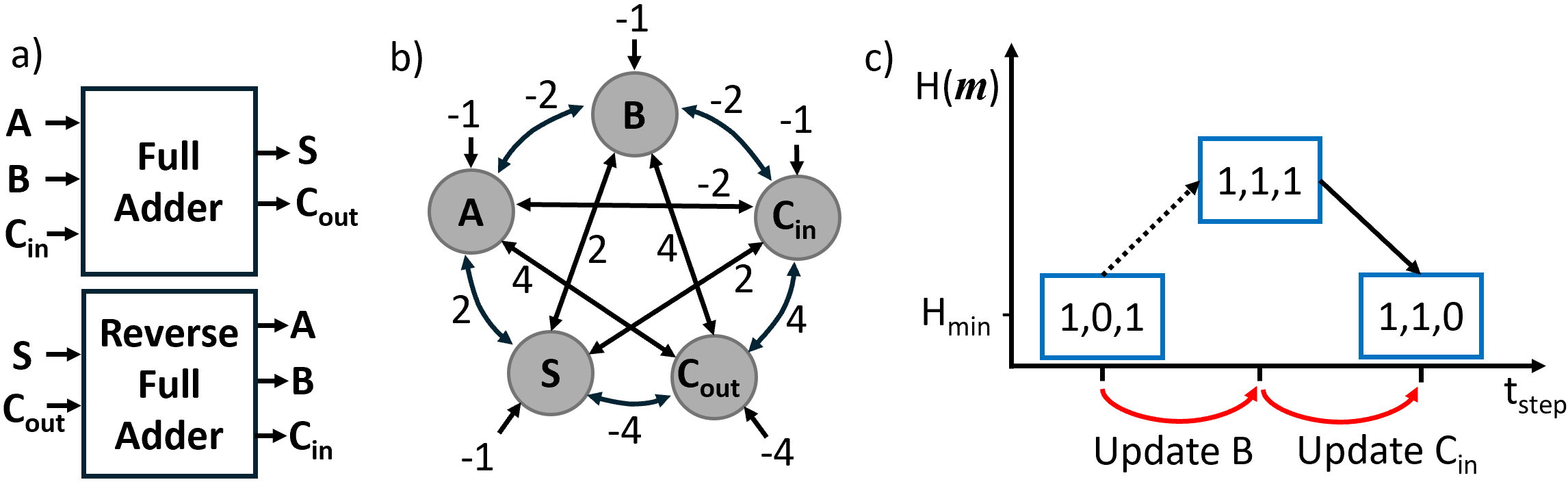}
    \caption{\textbf{Full-adder.} a) Typical circuit diagram. b) Probabilistic formulation, with weights and biases in binary form. c) Example input (A,B,C$_{\textrm{in}}$) state evolution in reverse mode, initial condition $= \{1,0,1\}$, outputs clamped to \{S,C$_{\textrm{out}}$\}=\{0,1\}. Dotted lines are energetically unfavorable flips (with $I_i=\pm1$). Solid lines are energetically favorable flips. }
    \label{fig:FA}
\end{figure*}

The full-adder (FA) is a 5-P-Bit gate shown in Fig.~\ref{fig:FA}a) with weights and biases given in Fig.~\ref{fig:FA}b). In forward mode, the three input bits $\{A,B,C_{in}\}$ are clamped to input values, and the system self-organizes to produce the 2-bit sum $\{S,C_{out}\}$. In the forward mode of operation, there is a single entry in the truth table corresponding to each input. 

In reverse mode, clamping $\{S,C_{out}\}=\{0,0\}$ ($\{1,1\}$) results in a single state $\{A, B,C_{in}\}=\{0,0,0\}$ ($\{1,1,1\})$ at the energy minimum and no randomness is needed to drive the network to these states.

There are multiple energy-minimum states when clamping $\{S,C_{out}\}=\{0,1\}$ or $\{1,0\}$  For $\{S,C_{out}\}=\{0,1\}$, these are $\{A,B,C_{in}\}=\{1,1,0\}$, $\{1,0,1\}$, and $\{0,1,1\}$, which should be visited with equal probability.  Unlike the AND P-Bit network, each energy-degenerate FA state resides in an isolated energy minimum.  Transitions between them require an energy-increasing P-Bit update provided by the randomness.

An example sequence of the P-Bit updates with their corresponding energy is shown in Fig.~\ref{fig:FA}c). Without an energetically unfavorable flip, the input P-Bits will remain in a single state. Without an energy-increasing update (due to the randomness), the state will stay in $\{1,0,1\}$ forever. This illustrates the need for probabilistic updates.

\begin{figure}
    \centering
    \includegraphics[width=0.98\linewidth]{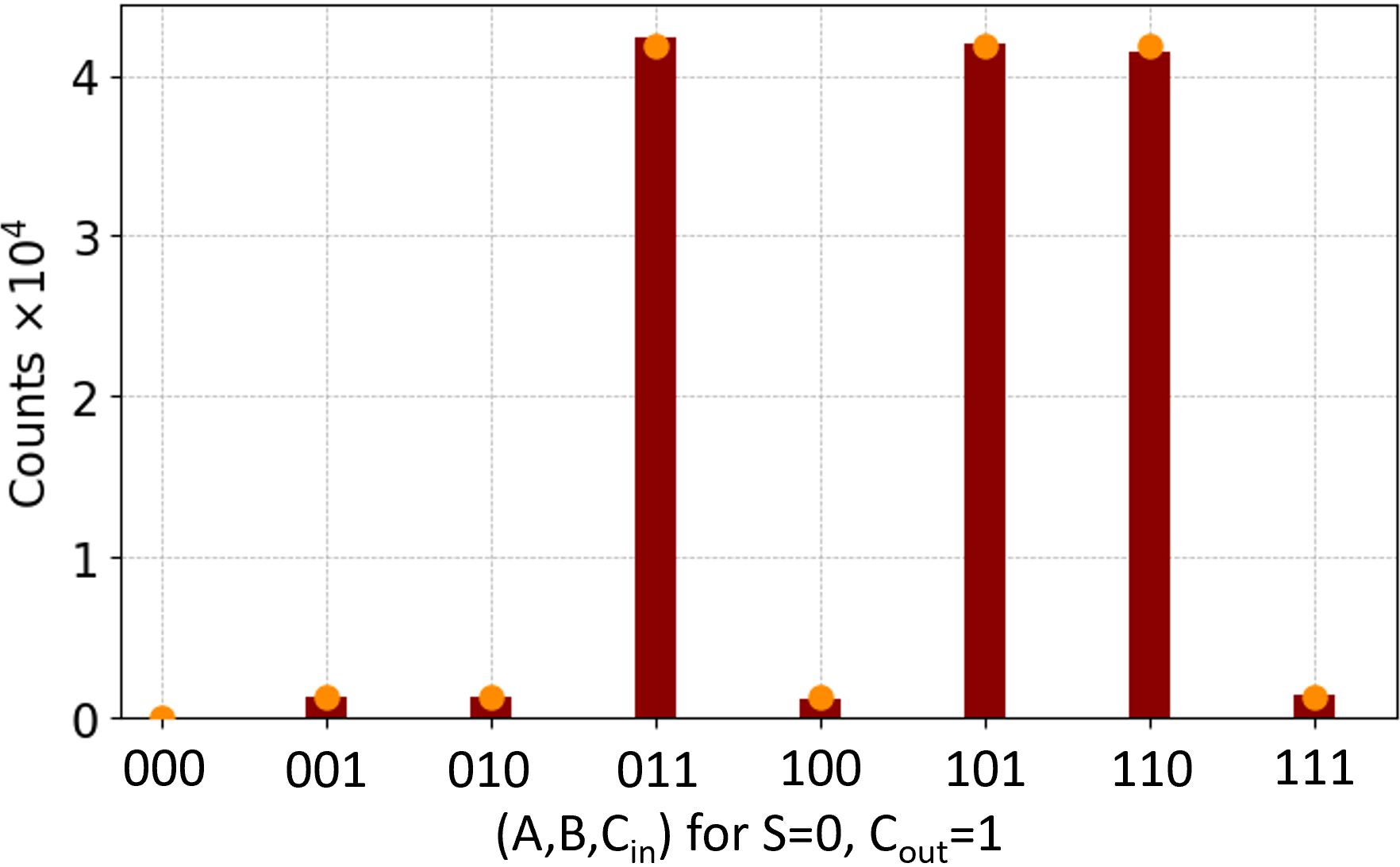}
    \caption{\textbf{Full adder experiments.} Frequency diagram for the FA operating in reverse mode with $S=0$ and $C_{\textrm{out}}=1$ for the 131,072 full system updates. The orange dots show the Bolztmann probabilities given by Eq.~(\ref{Eq:probs}).}
    \label{fig:FAdata}
\end{figure}

Figure~\ref{fig:FAdata} shows the experimental results for the probabilistic FA gate operating in reverse mode with $\{S,C_{out}\}=\{0,1\}$. The circuit visits each global energy minimum with nearly equal frequency. We update $A$, $B$, and $C_{in}$ sequentially at 350 MHz on the positive edges of the multiple clock outputs, each phase-shifted by $90^{\circ}$. We measure the state on the positive edge of a third, equally phase-shifted clock using the ILA. For the cases when $|I_i|=1$, we use $R=5$ to generate $b$ in Eq.~\textcolor{black}{(\ref{eq:b_final})}, which is equivalent to $\beta=1.717$ by comparing the flip probability of $I_i=\pm1$ cases. We also give the expected number of counts at each state given by the Boltzmann probability distribution from Eq.~(\ref{Eq:probs}). \textcolor{black}{We find that we obtain the expected Boltzmann probability distributions for a wide range of the bias parameter $b$, and hence we hold it constant.} 

The input P-Bit state is initialized to $\{A,B,C_{in}\} = \{0,0,0\}$. It immediately transitions to an allowed state after updating the first P-Bit. The $\{0,0,0\}$ state is never revisited because it requires updates not supported by Eq.~\textcolor{black}{(\ref{eq:b_final})}. Similarly, for $S=1$ and $C_{\textrm{out}}=0$, $\{A,B,C_{in}\}= \{1,1,1\}$ is never visited. 

For the FA, we perform 1.05$\times10^9$ P-Bit FPS.  We do not maximize the clock frequency for this small probabilistic gate, and the lower clock frequency used here is also likely due to the somewhat larger ILA required to collect data from this network.  The on-chip resources for this network are 27 LUTs and 36 FFs, not including the ILA for data collection, which consumes an additional 1058 LUTs and 1848 FFs.

\section{Factoring semiprimes\label{Sec:multipliers}}

We use conventional binary long multiplication to construct a P-Bit network that can be run in reverse mode to factorize semiprimes. The binary partial products are found using AND gates, which are summed using FAs as described in Appendix~\ref{Appendix:Multiplier}.

We construct the full-circuit interactions and biases by combining the  \textbf{J} matrix and \textbf{h} vector for AND and FA gates shown in Fig.~\ref{fig:AND}b) and \ref{fig:FA}b). We sparsify the circuit following the procedure described in Aadit \textit{et al.} \cite{Aadit2022} and summarized in Appendix~\ref{Appendix:Multiplier}.  For a sparsified $k-\mathrm{bit} \times k-\mathrm{bit}$ multiplier with a maximum of five connected neighbors ($M\leq5$), there are $k^2$ AND gates, $k(k-1)$ FAs, and $5k(k-1)-3k$ COPY gates between them. This expression includes COPY gates between FAs and between FAs and ANDs. Each input bit is distributed to $k$ AND gates. We add $r=1\rightarrow log_{M-1}(k)$ layers of P-Bits to sparsify each input bit, as shown in Fig.~\ref{fig:ANDhierarchy} of the Appendix. Layer $l_0$ consists of the $k$ input P-Bits distributed to AND gates. Each $l_r$ adds an additional 
\begin{equation} \label{eq:layers}
    l_r=\textrm{ceil}\left[\frac{l_{r-1}}{M-1}\right],
\end{equation}
P-Bits, where the ceil function rounds its argument to the next highest integer. Sparsifying the input P-Bit distribution also adds $l_r$ additional COPY gates per layer, excluding the top layer, which will always be $l_r=1$. The P-Bit at the top layer is the one we measure when running the multiplier in reverse mode. We also require $R$ random bits for each P-Bit with an instance of $|I_i|=1$, and one random bit for each P-Bit with an instance of $I_i=0$. We divide these by color, and generate $\textrm{RNG}_{\textrm{total},col}/46$ 46-bit LFSRs for color $col$. Random bits are drawn at random from the entire list of $\textrm{RNG}_{\textrm{total},col}$ random bits without replacement to form $b$ in Eq.~\textcolor{black}{(\ref{eq:b_final})}.

Below, we consider multipliers from 3-bit $\times$ 3-bit up to 16-bit $\times$ 16-bit. We use the ILA for data collection as we did in Sec.~\ref{Sec:LogicGates}. We add the Xilinx virtual input-output (VIO) IP block to change products and reset the system, resulting in a general-purpose factorizer. We find the best time-to-solution when using $R=3$ ($R=4$) for the biased random number generator for $k<16$ ($k=16$). \textcolor{black}{We selected $R$ by minimizing the time-to-solution using a grid search. The optimal $R$ presents the best balance between rapidly exploring the solution space (requiring more noise, better with smaller $R$) and finding energy minima (requiring less noise, better with larger $R$). We use a larger $R=5$ for the $3\times3$ multiplier analysis to increase the ratio of probability to be in allowed states to that of unallowed states.}

\subsection{Small multiplier\label{subsec:smallMult}}

The optimization energy surface becomes increasingly complicated as the problem scale increases, such as for a multiplier composed from many smaller gates.  This requires additional search time to identify the global minimum.  We clamp the output bits to the semiprime to be factored and randomly initialize the remaining network P-Bits.  We then execute 1,000 full system updates and read the input P-Bits corresponding to the two prime factor candidates. \textcolor{black}{This value is unrelated to the number of updates typically required to find a solution; For any more than 1,000 full system updates, state visitation statistics change little, indicating 1,000 updates are enough for the system to settle into deeper energy minima.} We repeat this sequence 131,072 times.

A full system update involves updating all the P-Bits that share a color simultaneously and updating the colors sequentially. Each color updates on a separate phase-shifted clock edge. The LFSRs that generate the RNGs for one color are updated simultaneously on the previous color's clock edge. We use the mixed-mode clock manager (MMCM) block on the FPGA to realize parallel 110 MHz clocks, using the 100 MHz low-voltage differential signaling (LVDS) reference clock as input. The result is a set of highly stable, synchronized, low-phase-noise clocks that result in predictable system behavior. We add one additional equally phase-shifted clock (from the MMCM) that enables state readout after each full system update. We consider P-Bit multiplier networks with $M=5$, so we have six 110 MHz clocks separated by a $60^{\circ}$ phase shift between each. Every P-Bit is updated in a single clock cycle.

\begin{figure}
    \centering
    \includegraphics[width=\linewidth]{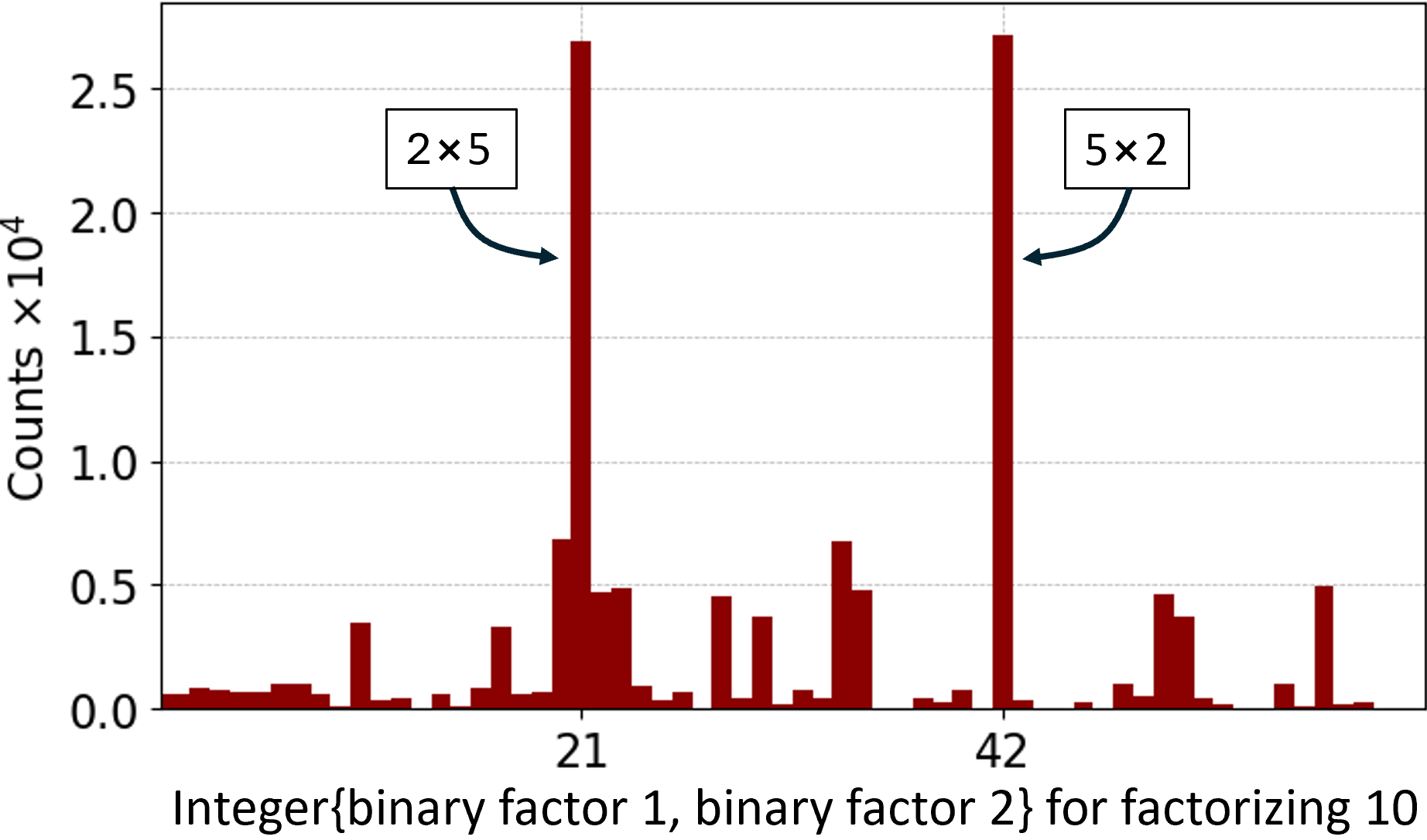}
    \caption{\textbf{Small multiplier experiments.} Frequency diagram for the input bits of a 3$\times$3 multiplier run in reverse mode when the output bits are set to $10$ in base 10. }
    \label{fig:3x3Mult}
\end{figure}

Figure~\ref{fig:3x3Mult} shows the `input' P-Bits (factors) state statistics for the output clamped to $10_{10} = 001010_{2}$, where the subscript denotes the number base. \textcolor{black}{Scaling the vertical axis in Fig. 7 by the total number of samples transforms it into the state probability distribution. We note that the input state frequency statistics disregard the states of the auxiliary P-Bits.}  The correct solutions are $2_{10}\times 5_{10}$ = $010_{2} \times 101_2$ or $5_{10}\times 2_{10}$ = $101_{2} \times 010_2$.  The horizontal axis indicates the factors in the format of the first factor concatenated by the second factor in binary format, then this number is converted to base 10.  Using this convention, solutions are $010101_2 = 21_{10}$ and $101010_2 = 42_{10}$. These solutions correspond to the tallest two bars.

The smaller peaks in the plot have contributions from two cases: either the system had an energy-increasing update just before the sample, or the state is in a local minimum. Sampling after one thousand updates minimizes contributions from local minima to the state population histogram, giving us a clearer picture of the energy landscape. Frequently visited non-solution states correspond to deeper energy minima that require more simultaneous energetically unfavorable flips to escape. 

For the $3\times3$ multiplier, we perform $5.94\times10^9$ P-Bit FPS. We calculate this value by taking the total number of P-Bits in the network (63), subtracting those that are clamped [outputs (6) and the zero-input P-Bits to the FAs (3), see Appendix~\ref{Appendix:Multiplier}], and multiplying the result by clock frequency. We do not maximize the clock frequency for this smaller multiplier, instead keeping it the same for all $k$. The lowest clock frequency is for $k=16$ discussed below. The on-chip resources for this network are 131 LUTs and 345 FFs, not including the ILA for data collection or the VIO for changing memory cells for the semiprimes, which together consume an additional 1,597 LUTs and 2,693 FFs.

\subsection{Larger multipliers\label{subsec:largerMults}}

We study the time-to-solution for multipliers running in reverse mode up to $k=16$\textcolor{black}{, showing significant ($\gtrsim1$ order-of-magnitude) improvements over similar approaches. Similarly, we show a $>1$ order-of-magnitude decrease in the number of on-chip resources required to construct multiplier circuits.} For each, we keep the biased random number generator on all the time for cases of $I_i=\pm1$, with no annealing schedule.  When the correct solution is found, some P-Bits in the network have $I_i=\pm1$, and hence the system is eventually driven out of the global minimum once there is a bit flip that increases the energy of the state.

For this reason, we add an oracle that checks whether factors match the product state after each full system update using on-chip dedicated multipliers on the FPGA. Once a solution is found, the circuit reports the number of full system updates, sets the unclamped P-Bits to random initial conditions, and restarts the search.  We typically repeat this process 1,024 times. We perform fewer trials for the 16-bit $\times$ 16-bit multiplier because of the longer time-to-solution, but we never use fewer than 13. We repeat this process for ten semiprimes for each multiplier size. The semiprimes are constructed by arbitrarily choosing $k$-bit prime integers, avoiding repetition. Repetition is unavoidable for $k=6$ because there are only seven prime integers in this case.

As we did for the smaller ($k=3$) multiplier in Sec.~\ref{subsec:smallMult}, we use $M=5$ for all multipliers here. We perform updates for all multipliers (all $k$) using 110 MHz clocks, identical to those we used for the $k=3$ multiplier. \textcolor{black}{Because the clock frequencies do not change, the data we present in terms of time-to-solution can be converted to iterations-to-solution by multiplying by 110 MHz.} For $k=6-14$, $R=3$ results in the most efficient time to solution owing to its rapid exploration of the solution space. For $k=16$, we find that $R=4$ minimizes time to solution.

The $k=16$ multiplier performs $2.29\times 10^{11}$ FPS (2,128 total P-Bits, 32 clamped to the product state and 16 clamped to zero for 2,080 P-Bits updated every clock cycle), with as many as $6.69\times10^{10}$ FPS for one color. As Aadit \emph{et al.} \cite{Aadit2022} find, the flips per second grow linearly with problem size. We estimate that they observe $4.33\times 10^{10}$ FPS in their $k=16$ multiplier based on the scaling linearity in combination with other provided data.  Our observed speedup comes from the increased clock rate.

\begin{figure}
    \centering
    \includegraphics[width=\linewidth]{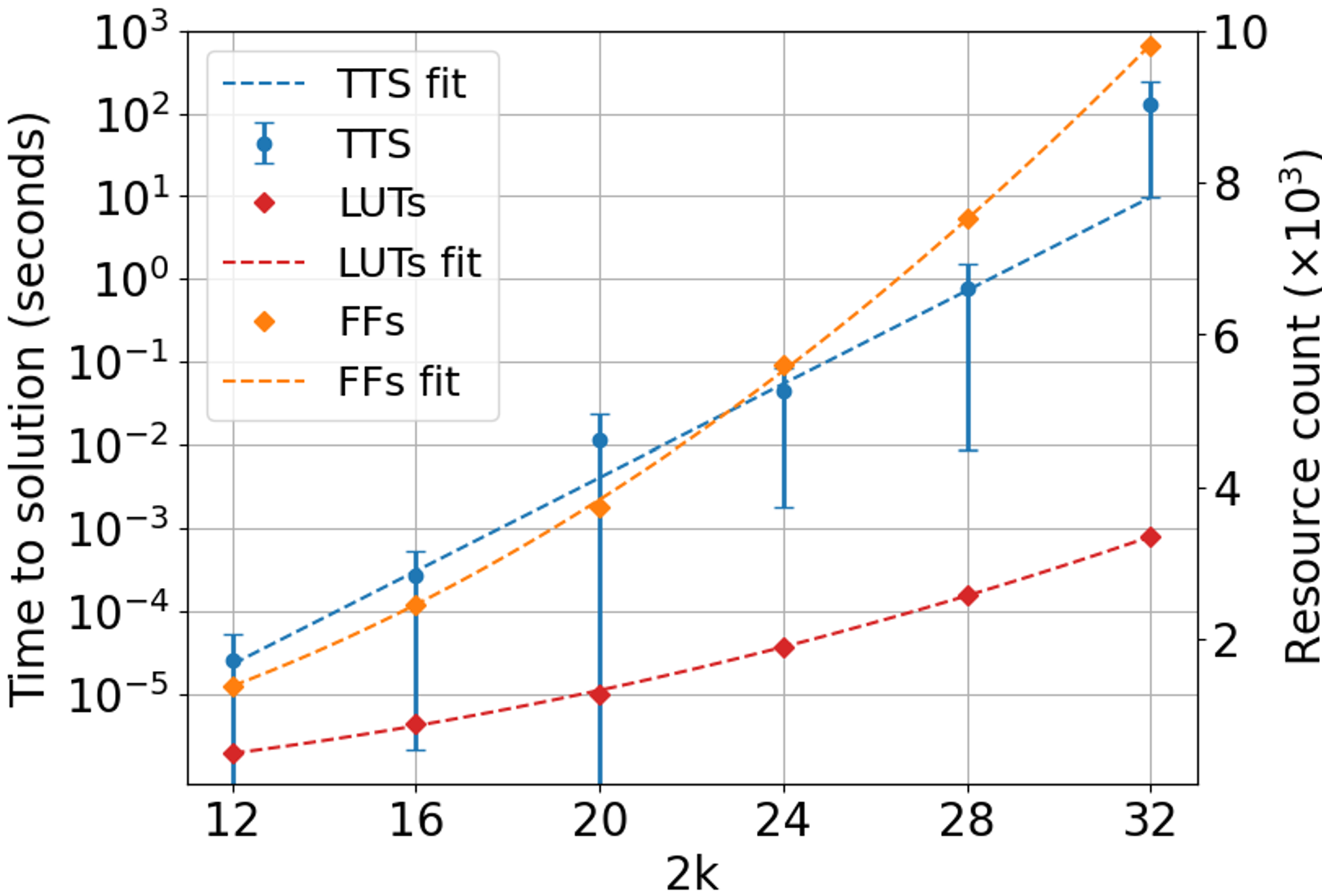}
    \caption{\textbf{Performance \textcolor{black}{and on-chip resources} of the multiplier circuits running in reverse mode.} The average time required to factorize different product sizes \textcolor{black}{(blue dots/dashed line), and the number of LUTs (red dots/dashed line) and FFs (orange dots/dashed line) of the circuit.}}
    \label{fig:MultiplierData}
\end{figure}

Figure~\ref{fig:MultiplierData} shows the scaling of the average time-to-solution (blue dots and line of best-fit) as a function of $2k$ (the number of bits in the semiprime) on a semi-logarithmic scale \textcolor{black}{(left vertical axis)}. The data is fit to a straight line, resulting in the expression of best-fit $(10^{-8}\textrm{ seconds})\times \textrm{exp}(2k/1.5)$. \textcolor{black}{Figure~\ref{fig:MultiplierData} also shows the total number of each on-chip resource used to realize each multiplier circuit plotted on a linear scale (right vertical axis). The scaling is quadratic, as expected from the quadratic scaling of the number of P-Bits with multiplier size. The line of best-fit for LUTs is $3.24\times(2k)^2+30$ and for FFs is $9.59\times(2k)^2+1$. The resource utilization data excludes resources used by the ILA for data collection and the VIO for changing semiprimes, but it includes the oracle for indicating that a solution has been found.}

The exponential scaling of the time-to-solution as a function of problem size demonstrates that the P-Bit network is not an efficient semiprime factorizer.  However, the observed scaling for our architecture and absolute time to solution are better than reported by Aadit \textit{et al.} \cite{Aadit2022}. They find a best-fit of $(10^{-7.17}\textrm{ seconds})\times\textrm{exp}(2k/1.26)$, which has a larger prefactor and a larger exponent compared to ours. They also report an average time to solution of over $10^3$ s for $k=16$, which is an order-of-magnitude larger than our result.

Our approach also uses fewer on-chip resources.  The sparsified architecture we use here is identical to that of Aadit \emph{et al.} \cite{Aadit2022}, but they use the P-Bit circuit shown in Fig.~\ref{fig:typicalCircuit}.  Unfortunately, they do not report their chip resource usage.  Table~\ref{tab:resources} compares our chip usage to a similar long-multiplication architecture but with a dense (unsparsified, $M\gg1$) connectivity from Onizawa \emph{et al.} \cite{Onizawa2021_DesignFramework}. Even though we increase the total P-Bits and the number of nonzero weights, we use over an order-of-magnitude fewer on-chip LUTs and a comparable number of FFs. We exclude ILA for data collection and VIO for altering the semiprime from the resource comparison. Together, these auxiliary resources consume an additional 4,558 LUTs and 4,875 FFs. Their unsparsified architecture does not allow for graph coloring, massively parallel P-Bit updates, and our approach of using small LUTs for the update rule and hence we expect that their time-to-solution will be substantially longer. 

\begin{table}[hbt]
\caption{\label{tab:resources}
\textbf{Chip area.} Resource usage between our 16$\times$16 bit factorizer and an unsparsified 16$\times$16 bit factorizer with the same underlying architecture.}
\begin{ruledtabular}
\begin{tabular}{ccc}
 Resource & Typical CMOS method\footnotemark[1]&Our Method \\
\hline
LUTs & 62,969 & 3,350\\
Flip-Flops & 9,393 & 9,798 \\

\end{tabular}
\end{ruledtabular}
\footnotetext[1]{From Onizawa \emph{et al.} \cite{Onizawa2021_DesignFramework}.}
\end{table}


\section{Conclusions\label{Sec:Conclusions}}

Our experiments demonstrate that new P-Bit update rules can greatly simplify the circuit used to simulate P-Bit network dynamics. We focus on running the network in the reverse model because standard CMOS logic can be used to find the solution in the forward mode.  We find that the time-to-solution and the on-chip resource usage is reduced substantially, sometimes by orders-of-magnitude, compared to the state-of-the-art reported in the recent literature.  Using an FPGA with our new update rule, a sparsified circuit, and graph coloring allows for massively parallel execution of the search.

It is not surprising that we obtain an exponential scaling of the time-to-solution for factoring semiprimes because we are performing a simulation of a P-Bit network using clocked logic.  The P-Bit network is effectively performing a multi-dimensional energy-minimizing search with occasional random bit flips. This is similar to other approaches to this problem, such as basin hopping \cite{Wales1997}, which also has an exponential scaling with problem size.  

Every multiplier we constructed and tested in Sec.~\ref{Sec:multipliers} is a general circuit with resettable outputs (the semiprime) when operating in reverse mode by loading different bit values into on-chip memory. Also, every $k$-bit$\times$ $k$-bit multiplier can be used for $(\leq k)$-bit $\times(\leq k)$-bit problems with similar time-to-solution scaling as in Fig.~\ref{fig:MultiplierData}.

Recent P-Bit research considers higher-order interactions \cite{Bashar2023,Bybee2023,Nikhar2024,He2024}.  That is, the Hamiltonian (\ref{Eq:H}) has terms that include cubic, quartic, etc. monomials of the spins.  The problem has a compact form, where many spins are grouped into a simplex structure, resulting in a simpler energy landscape.  However, there is a limit to the size of a simplex because we must store the interaction weights, and we require many logic bits on an FPGA to represent the higher-order spins.  Further research is needed to understand the trade-off of these approaches.

Our approach is not limited to integer factorization. We expect that many probabilistic hard combinatorial problem solvers would benefit from the simplified update rules, such as the Max-Cut or 3-SAT problems. Our simplifications can also be introduced to probabilistic/stochastic machine learning algorithms for fast and accurate training \cite{Kaiser2020}.

We close by mentioning that our new update rule may suggest new analog approaches to creating P-Bit networks and potentially better scaling with problem size. Much of the P-Bit literature focuses on P-Bit networks realized with \textcolor{black}{magnetic tunnel junctions} because they have a graded response with input current and have thermal noise \textcolor{black}{\cite{Sutton2020,Borders2019,Si2024}}. Because we no longer need a graded response directly in the P-Bit structure [the hyperbolic tangent function inside the sgn function of Eq.~(\ref{Eq:I})], other physical devices might be good P-Bit candidates, such as using autonomous logic (CMOS logic operated without a clock) \cite{Rosin2013}.

\begin{acknowledgments}
We gratefully acknowledge the financial support of the Air Force Research
Lab (AFRL) under Agreement FA8650-19-1-1741 and conversation about this research with the members of the Ohio State Univesity Center for Enabling Cyber Defense in Analog and Mixed Signal Domain (CYAN).
\end{acknowledgments}

The code and data that support the findings of this article are openly available \cite{Code, Data}

\appendix

\begin{figure}[htb]
    \centering
    \includegraphics[width=\linewidth]{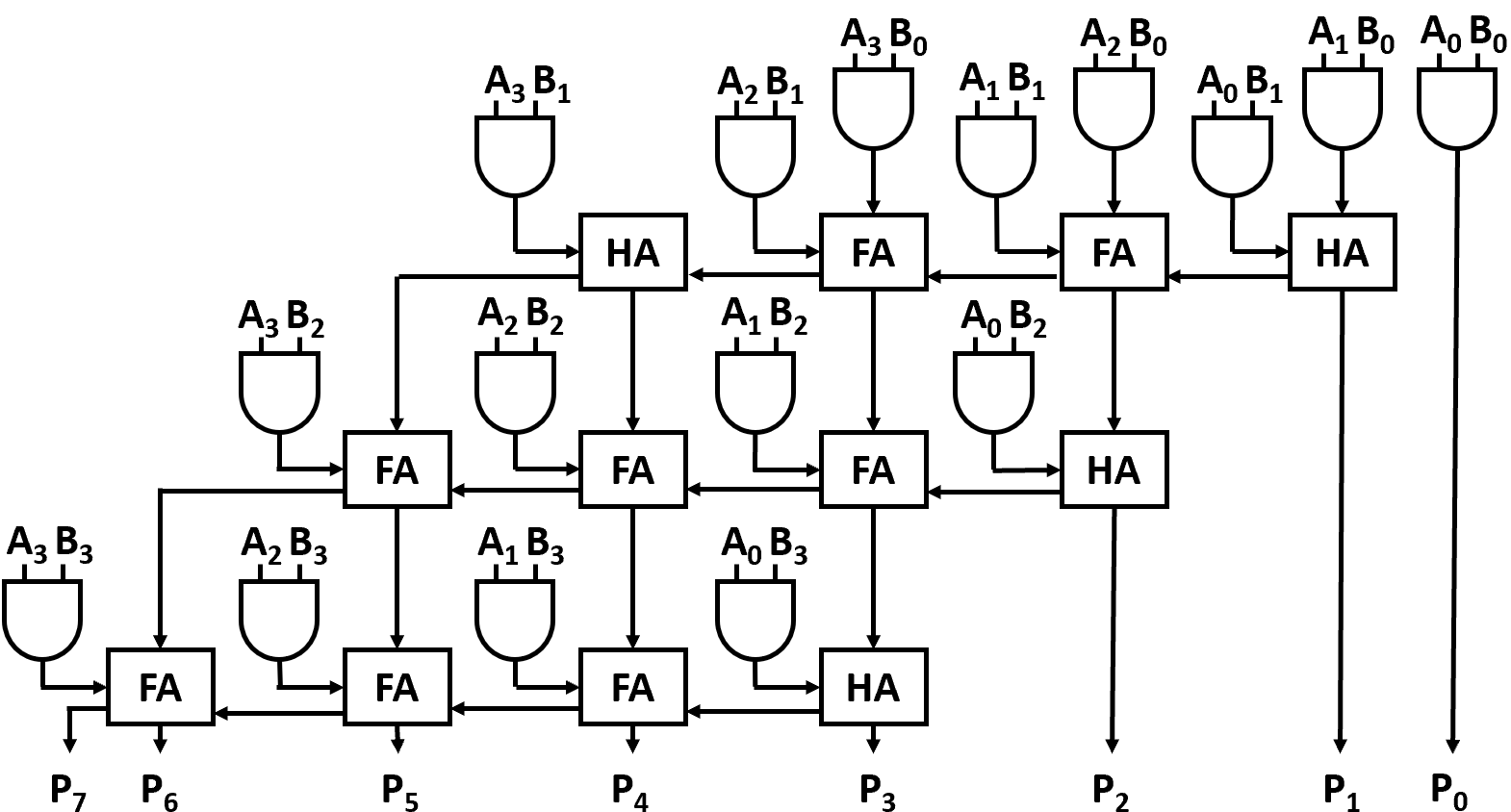}
    \caption{\textbf{Binary mutliplier.} Conventional 4-bit $\times$ 4-bit multiplier circuit. Input factors $A$ (most significant bit A$_3$, least significant bit A$_0$) and $B$, output product $C$. Each line and each input/output bit represents a P-Bit in the system.}
    \label{fig:ConventionalMult}
\end{figure}

\section{Multiplier structure \label{Appendix:Multiplier}}
As mentioned in the main text, we use a binary long-multiplication architecture, with an example of a 4-bit $\times$ 4-bit multiplier shown in Fig.~\ref{fig:ConventionalMult}. The three fundamental gates are the AND, Half Adder (HA), and FA. To simplify the software that automatically generates \textbf{J} and \textbf{h} for larger multipliers, we substitute the HA with FA and clamp the input carry bit to zero, which produces the same function.  These bits are ultimately removed by the Xilinx circuit compiler/optimzier (Vivado system) because they do not contribute to the system.

\begin{figure}[htb]
    \centering
    \includegraphics[width=\linewidth]{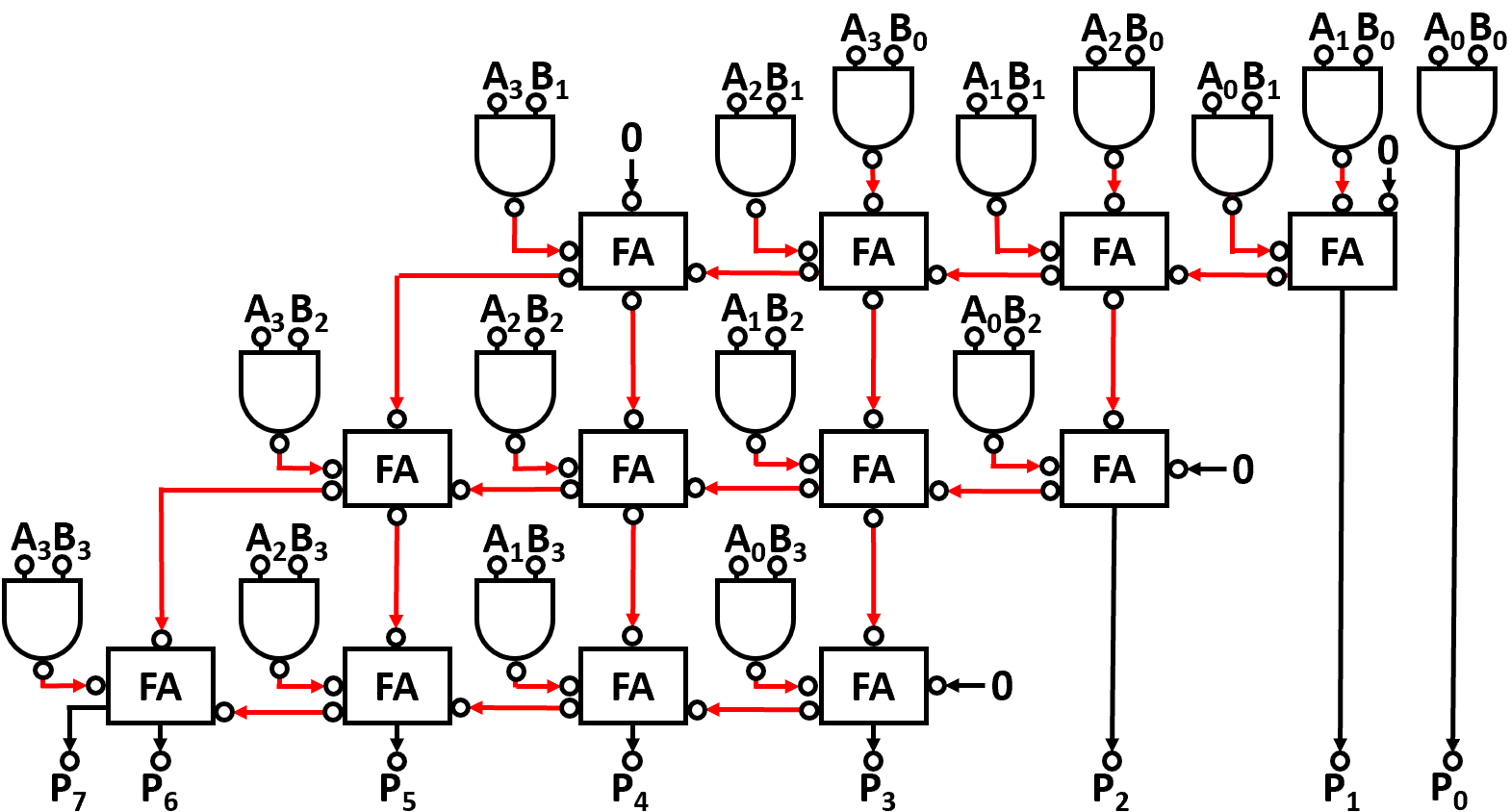}
    \caption{\textbf{Sparsified binary multiplier.} A sparsified 4-bit $\times$ 4-bit multiplier, where P-Bits are black open circles and COPY gates are red lines. We replaced HA gates with FA gates with $C_{in}$ clamped to $0$.}
    \label{fig:SparseMult}
\end{figure}

To sparsify the circuits, we add COPY gates between the AND and FA gates, and between each of the FAs as shown in Fig.~\ref{fig:SparseMult}. We use a maximum of five nearest neighbors for each P-bit ($M\geq5$), which aligns well with splitting P-Bits that act as a bridge between FA gates. 

\begin{figure}[htb]
    \centering
    \includegraphics[width=\linewidth]{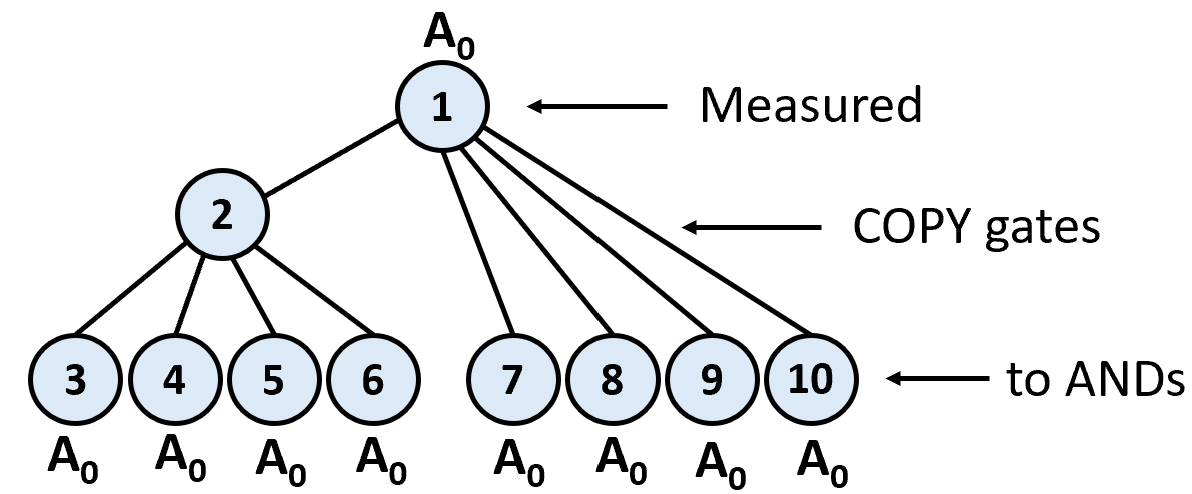}
    \caption{\textbf{Sparsifying fanout of the input P-Bits to the network.} Sparsification of a input P-Bit $A_0$ in an 8-bit $\times$ 8-bit multiplier. Lines are COPY gates and circles are P-Bits.}
    \label{fig:ANDhierarchy}
\end{figure}

The remaining dense part of the circuit is fanning the input P-Bits to the AND gates: Each input P-Bit must be distributed to $k$ AND gates.  To reduce the number of connections, we use a hierarchy of COPY gates. Figure~\ref{fig:ANDhierarchy} shows an example of this hierarchy for an 8-bit $\times$ 8-bit multiplier. In this case, we add 9 auxiliary P-Bits to reduce the number of connections to 5, starting from the initial configuration with 25 connections. 

\bibliography{ProbComputingABN.bib}

\end{document}